\documentclass[12pt,aps,a4paper,preprint, showpacs, showkeys, superscriptaddress]{revtex4}
\usepackage{graphicx}
\usepackage{comment}
\usepackage[breaklinks=true,colorlinks=true,linkcolor=blue,urlcolor=blue,citecolor=blue]{hyperref}
\usepackage{amsmath}

\begin{document}

\title[Radio frequency dressed potentials]{On trapping geometry for cold atoms in a radio-frequency (rf) dressed potential}

\author{Sourabh Sarkar}
\affiliation{Laser Physics Applications Division, Raja Ramanna Centre for Advanced Technology, Indore, 452013, India}
\affiliation{Homi Bhabha National Institute, Training School Complex, Anushakti Nagar, Mumbai, 400094, India}

\author{S. P. Ram}
\email{spram@rrcat.gov.in}
\affiliation{Laser Physics Applications Division, Raja Ramanna Centre for Advanced Technology, Indore, 452013, India}

\author{Kavish Bhardwaj}
\affiliation{Laser Physics Applications Division, Raja Ramanna Centre for Advanced Technology, Indore, 452013, India}

\author{Gunjan Verma}
\affiliation{Laser Physics Applications Division, Raja Ramanna Centre for Advanced Technology, Indore, 452013, India}

\author{V. B. Tiwari}
\affiliation{Laser Physics Applications Division, Raja Ramanna Centre for Advanced Technology, Indore, 452013, India}
\affiliation{Homi Bhabha National Institute, Training School Complex, Anushakti Nagar, Mumbai, 400094, India}

\author{S. R. Mishra}
\affiliation{Laser Physics Applications Division, Raja Ramanna Centre for Advanced Technology, Indore, 452013, India}
\affiliation{Homi Bhabha National Institute, Training School Complex, Anushakti Nagar, Mumbai, 400094, India}

\begin{abstract}
We have investigated the atom trapping geometry for trapping of $^{87}{Rb}$ atoms in a radio-frequency (rf) dressed potential generated after superposing a strong linearly polarized rf-field on a static magnetic trap. For this, laser cooled atoms in a magneto-optical trap (MOT) in an ultra-high vacuum (UHV) chamber (pressure $\sim$ 1.5 $\times$ $10^{-10}$ Torr) were trapped in a quadrupole magnetic trap and evaporatively cooled before transferring them to the rf-dressed potential. The experimentally observed hollow shell type atom trapping geometry has been explained by theoretical modelling of the trapping potential. 
\end{abstract}


\vspace{2pc}
\maketitle

\section{\label{sec:Intro}Introduction} Cooling and trapping of atoms give us a new and unique way to look into the quantum world of atoms. Since the inception of laser cooling and trapping of atoms, the physics of cold atoms spread its horizon to explore many fundamental phenomena such as Bose-Einstein condensation (BEC) \cite{Davis1995, Pethick_BEC}, complex phase transition \cite{Greiner_2002,Sunami_BKT}, tunnelling \cite{Albiez2005}, etc. Besides that, now a days, many practical applications of cold atoms are emerging in precision measurements of time \cite{Ludlow_Clock} , gravity \cite{Peters2001, OL_Gravity_Poli,Fountain_Gravi_Abend}, rotation \cite{Gus_1997,Area_Enclosed_Wu,Ragole_Int_Atom_Interferometry}, electric \cite{Electric_Field_Sensing_Carter} and magnetic fields \cite{Yuval_Magnetometer}. The cold atoms are also considered important to develop qubits for quantum information \cite{QIP_Margado} applications. All this is possible because of several new developments have taken place in the past several years to control and manipulate atoms using different techniques.  

 The technique of radio frequency (rf) dressing of atoms, first proposed by Zobay and Garraway \cite{Zobay2001}, is one such technique which provides unique opportunity to tailor the trapping geometries for atoms using rf-fields and static (i.e. DC) magnetic fields \cite{H_Perrin_2017,Chakraborty2014, Sarkar2021}. A good control on atom trapping geometries can be achieved in rf-dressed potential traps by changing magnetic field and rf-field parameters. This has been earlier investigated by several researchers \cite{ColombeY.2004, Morizot2006, Chakraborty2016, Schumm2005, Sherlock2011, Easwaran2010} which includes experimental demonstration of various trapping geometries such as ring trap \cite{Morizot2006, Sherlock2011, Heathcote2008}, double-well  trap geometry \cite{Schumm2005,T_L_Hartre_Multi,M_White_2006,Gildemeister2010}, etc. The ring trap is an ideal candidate to study the superfluidity \cite{Ramanathan2011, Heathcote2008, Gildemeister2012} and rotation of cold atom cloud \cite{Gildemeister2012,Pandey2019, Sourabh2021}. The double-well is useful for atom interferometry  \cite{Schumm2005}, tunnelling \cite{Albiez2005}, species selective confinement \cite{Bentine_2017}, etc. 
 
In this paper, we have investigated the trapping geometry of $^{87}$Rb atoms in a rf-dressed potentials formed by a static quadrupole magnetic trap and a strong rf-field. For this, the $^{87}$Rb atoms from a magneto-optical trap (MOT) in an ultra-high vacuum (UHV) chamber (pressure $\sim$ 1.5 $\times$ $10^{-10}$ Torr) were first trapped in a quadrupole magnetic trap. The atoms were then cooled by rf-evaporative cooling method to bring them to a lower temperature suitable for transferring them to the rf-dressed potential. With a linearly polarized rf-field, the experimentally observed trapping geometry was a hollow shell-like structure. The observed atom trapping geometry has been explained on the basis of theoretical modelling of the trapping potential. This type of atom trapping geometry may have potential applications in various studies such as low dimensional Bose-Einstein condensation (BEC) \cite{Merloti2013}, superfluidity \cite{Ramanathan2011, Heathcote2008}, atom interferometry \cite{Navez_2016}, etc.

This article is organized as follows in section \ref{sec:Theory} we have briefly described the theoretical formalism and understanding of rf-dressing. Section \ref{sec:expt} has been dedicated for the description of our experimental set-up, methods and experimental techniques. In section \ref{sec:Results} we have discussed our experimental results and finally we conclude in section \ref{sec:Conclusions}.

\section{\label{sec:Theory}Theoretical background}
A static or DC magnetic field which is inhomogeneous in space provides trapping potential for atoms via interaction with hyperfine spin. There are varieties of magnetic traps \cite{Pethick_BEC} that can be used for magnetic trapping of atoms. Here we consider the simplest DC magnetic trap which is known as quadrupole magnetic trap. The atoms are assumed to be in a trappable Zeeman hyperfine state ($|F, m_{F}\rangle$). When a strong rf-field at a particular frequency interacts with these atoms, the energy levels of atoms are modified and the atoms experience the dressed state potential. This potential is called RF-dressed potential \cite{Lesanovsky_PRA_2006, Chakraborty2014}, can be tailored by changing the strength of trapping DC magnetic field as well as by changing the magnitude, frequency and polarization of rf-field,  

Here we consider the DC magnetic trap as a quadrupole trap whose magnetic field is described as, 

\begin{equation}
\label{Static_Field}
\mathbf{B_{S}(r)} = B_{q} (x\mathbf{\hat{x}} + y\mathbf{\hat{y}} -2z\mathbf{\hat{z}}),
\end{equation}
where $B_{q} $ is the radial ($x-y$) magnetic field gradient of quadrupole magnetic trap.

The potential energy of an atom, in a Zeeman hyperfine state $|F,m_{F}\rangle$, trapped in this trap at any point (x, y,z) can be written as, 
\begin{equation}
\label{Static_Potential}
V_{m_{F}} = g_{F}m_{F}\mu_{B}B_{q} \sqrt{x^2 + y^2 + 4z^{2}},
\end{equation}
where $m_{F}$ is the Zeeman hyperfine quantum number corresponding to the hyperfine state with angular momentum F, $g_{F}$ is the Lande's g factor for $|F,m_{F}\rangle$ state and $\mu_{B}$ is the Bohr magneton.

Now when a rf-field of frequency $\omega_{rf}$ and amplitude $B_{rf}$ is applied on atoms trapped in the quadrupole magnetic trap, the modified potential energy $V_{D}$ including the gravitational potential can be written as \cite{Chakraborty2016},
\begin{equation}
\label{Dressed_Potential}
V_{D} = m_{F}\hbar\sqrt{{\Delta(r)}^2 + \Omega_{R}(r)^{2}} + mgz, 
\end{equation}
where $\Delta(r)$ is the detuning of rf-field with the local Larmor frequency and $\Omega_{R}(r)$ is the inhomogeneous Rabi frequency, $m$ is the mass of $^{87}Rb$ atom, and g is the earth's gravitational acceleration. 

The expressions for detuning and Rabi frequency are given as \cite{Chakraborty2014, Sarkar2021},
\begin{equation}
\label{Detuning}
\Delta (r) = \left|{\frac{g_{F}\mu_{B}B_{q} \sqrt{(x^2 + y^2 + 4z^2)}}{\hbar} - \omega_{rf}}\right|,
\end{equation}
and
\begin{multline} \label{Rabi Frequency}
{\mid\Omega_{R}(r)\mid}^{2}  = {\Big(\frac{{{g}_{F}{\mu}_{B}}}{2\hbar}\Big)}^{2}\Bigg[\frac{4{z}^2}{x^2 + y^2 + 4{z}^2}\Bigg(\frac{{\Big(B_{rf}^{x}\Big)}^2 {x}^2 + {\Big(B_{rf}^{y}\Big)}^{2} {y}^{2} }{{x}^{2} + {y}^{2}}\Bigg) + \Bigg(\frac{{\Big(B_{rf}^{x}\Big)}^{2} {y}^{2} + {\Big(B_{rf}^{y}\Big)}^{2} {x}^{2}}{{x}^{2} + {y}^{2}}\Bigg) \\
+ {\Big(B_{rf}^{z}\Big)}^{2}\Bigg(\frac{{x}^{2} + {y}^{2}}{{x}^{2} + {y}^{2} + 4{z}^{2}}\Bigg) - \frac{2 {B}_{rf}^{x} {B}_{rf}^{y} x y \cos({\phi}_{rf}^{y})}{{x}^{2} + {y}^{2} +4{z}^{2}} + \frac{4 {B}_{rf}^{x} {B}_{rf}^{y} z \sin({\phi}_{rf}^{y})}{\sqrt{{x}^{2} + {y}^{2} +4{z}^{2}}} + \frac{4 {B}_{rf}^{y} {B}_{rf}^{z} y z \cos({\phi}_{rf}^{y} - {\phi}_{rf}^{z})}{{x}^{2} + {y}^{2} +4{z}^{2}} \\
+ \frac{2 {B}_{rf}^{y} {B}_{rf}^{z} x \sin({\phi}_{rf}^{y} - {\phi}_{rf}^{z})}{\sqrt{{x}^{2} + {y}^{2} +4{z}^{2}}} + \frac{4 {B}_{rf}^{z} {B}_{rf}^{x} x z \cos({\phi}_{rf}^{z})}{{x}^{2} + {y}^{2} +4{z}^{2}} + \frac{2 {B}_{rf}^{z} {B}_{rf}^{x} y \sin({\phi}_{rf}^{z})}{\sqrt{{x}^{2} + {y}^{2} +4{z}^{2}}}\Bigg],
\end{multline}
where the symbols $B_{rf}^{x},B_{rf}^{y},B_{rf}^{z}$ denote the magnitudes of the applied rf-field along $x$, $y$ and $z$-directions respectively. The symbols $\phi_{rf}^{y}$ and $\phi_{rf}^{z}$ are the phase values of the $y$- and $z$-components of the rf-field relative to the $x$-component of rf-field. From Eq. (\ref{Dressed_Potential}), it is obvious that the rf-dressed potential $V_{D}$ is position dependent via detuning and Rabi frequency.\\

Now if we consider a special case where rf-field is linearly polarized along $x$-direction, the expression for Rabi frequency $\Omega_{R}(r)$ can be given by,

\begin{equation}
{\mid\Omega_{R}(r)\mid}^{2} = {\Bigg(\frac{{{g}_{F}{\mu}_{B}B_{rf}^{x}}}{2\hbar}\Bigg)}^{2}\Bigg[\frac{4{x}^{2}{z}^2}{({x}^2 + {y}^2)(x^2 + y^2 + 4{z}^2)} + \frac{y^2}{{x}^{2} + {y}^{2}}\Bigg].
\end{equation}
This expression can be used to understand the local variation in the rf-dressed trapping potential $V_{D}$ when a linearly polarized ($x$-direction) rf-field is used for rf-dressing.

\section{\label{sec:expt}Experimental Set-Up}

\begin{figure}
\includegraphics[width=0.98\textwidth,height = 0.25\textheight]{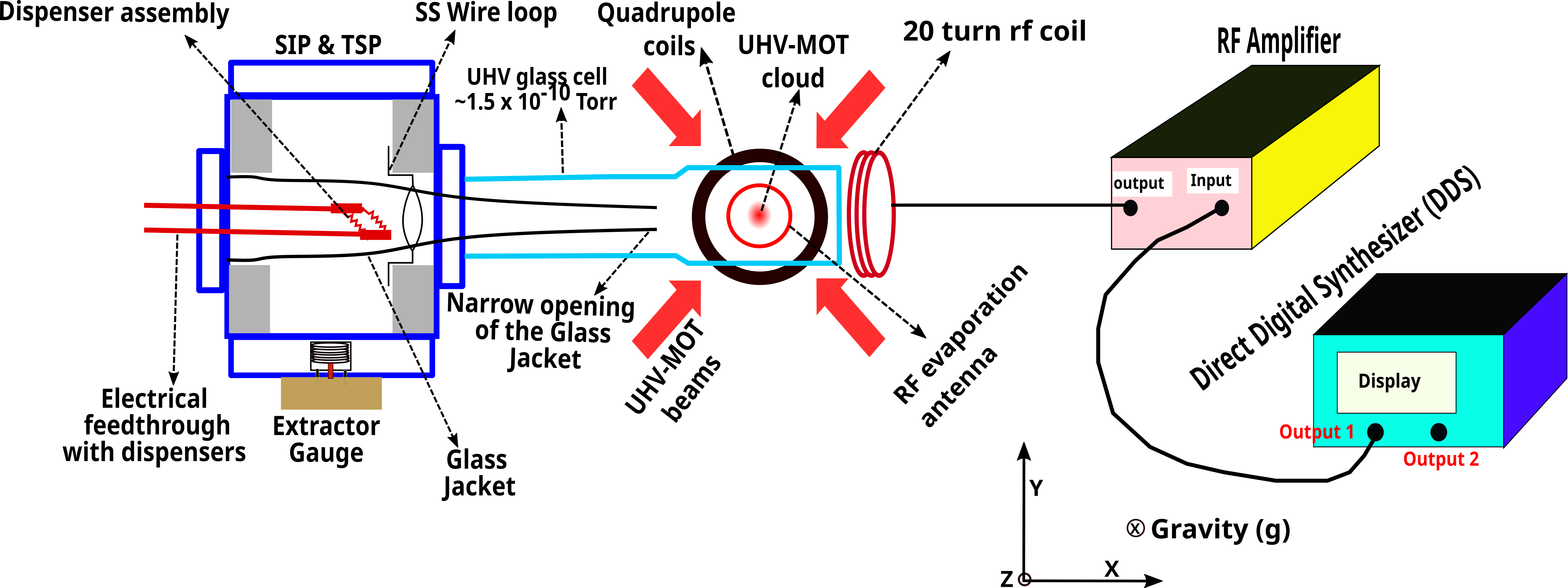}
\caption{\label{fig:Set_Up_Drawing}: The schematic of the experimental setup for rf-dressed atom trap.}
\end{figure}

The experiments have been performed in an in-house developed setup, where a single glass cell at ultra-high vacuum (UHV) pressure is used for performing all the sequences in the experiment from loading of a magneto-optical trap (MOT) to achieving a rf-dressed trap. The details of our setup has been given elsewhere \cite{Kavish_2022}. The schematic of the experimental set up has been shown in Fig. \ref{fig:Set_Up_Drawing}. The UHV glass cell was evacuated to a pressure level of $\sim$1.5 $\times$ $10^{-10}$ Torr. This UHV glass cell has its outer area of cross-section as 30 mm $\times$ 30 mm, and 70 mm length usable for optical beams. The glass cell wall thickness is 2.5 mm. The glass cell was attached to a six-way cube and kept horizontal in orientation.  

First of all, the magneto optical trap for $^{87}Rb$ atoms is prepared in the glass cell. For this, six independent cooling laser beams (shown as UHV-MOT beams in Fig. \ref{fig:Set_Up_Drawing}) are aligned on the glass cell. Each of these cooling beam was derived from a single laser (TA-Pro, Toptica, Germany), and each beam had a power of 20 mW at the glass cell entrance. A pair of repumping beam has been derived from another laser (DL-100, Toptica, Germany) and these beams were mixed with two cooling beams incident on the glass cell ($z$-directions beams). Each repumping beam has a power of around 10 mW. Both the lasers were respectively frequency locked to cooling and repumping transitions of $^{87}Rb$ and acousto-optic modulators (AOMs) were used to change the frequency as well as amplitude of the cooling and repumping laser beams as per requirement. For the MOT formation, the cooling beams are typically kept red-detuned by 2.5$\Gamma$ from the cooling transition ($5^{2}S_{1/2}$, $F = 2$ $\rightarrow$ $5^{2}P_{3/2}$, $F^{\prime} = 3$), where  $\Gamma = 2\pi \times 6.06$ MHz is the natural linewidth of the $^{87}Rb$ atoms cooling transition. The magnetic field gradient ($B_{q}$) of $\sim$5 G/cm is obtained for MOT from a pair of quadrupole coils operating with DC current of 1 A in anti-Helmholtz configuration. These coils, with higher currents, are also used for magnetic trapping in a subsequent stage of experiment. The Rb-source (dispensers) is kept inside a glass jacket which has a narrow opening near the centre of MOT (Fig. \ref{fig:Set_Up_Drawing}). This jacket facilitates to load MOT without degradation of UHV in the MOT region \cite{Kavish_2022}. In the MOT, we can have $\sim$1 $\times$ $10^8$ atoms $^{87}Rb$  at temperature of 250 $\mu$K. 

After loading the MOT, the optical molasses was implemented for a few milliseconds to reduce the temperature of atoms further. For molasses, the magnetic field for MOT was switched-off and cooling beam detuning was increased from -2.5$\Gamma$ to -8.0$\Gamma$. The temperature of the atom cloud after molasses was reached to $\sim$70 $\mu$K. After molasses, the atom cloud was prepared in a particular magnetic hyperfine state by implementing optical pumping. Optical pumping was done by exposing the atoms (in $F=2$ state) to a circularly polarized optical pulse ($\sim$ 500 $\mu$W of power and $\sim$ 500 $\mu$s of duration) in presence of a DC magnetic field of $\sim$ 2G. This prepares atoms in $|F = 2, m_{F} = 2\rangle$ for magnetic trapping. 

We used mechanical shutters in the path of the cooling beams which was necessary to fully stop any light going to the magnetic trap region. Usually, a leakage of laser beams through AOMs (when switched-off) is a primary source of light which affects the number of atoms in magnetic trap and trap life-time. For magnetic trapping, the optically pumped atom cloud was placed in quadrupole magnetic trap by switching-on the quardupole magnetic field with a lower gradient ($B_{q} = 35$ G/cm) which was held at that value for $\sim$ 100 ms. This step provided an initial loading of magnetic trap. After this initial loading, the gradient ($B_{q}$) was ramped up to a higher final value (e.g. $B_{q} = 126$ G/cm) slowly in 500 ms. At this final gradient, the atom cloud is held for further experiments. We observed that when final gradient was 126 G/cm, the atom cloud was observable in magnetic trap up to 11 s duration with a lifetime ($\frac{1}{e}$) of $\sim$ 4 s. In the magnetic trap, we have measured the number and temperature of the atom cloud, because it was important to know these parameters for subsequent rf-dressing experiment. Just after reaching the final gradient of 126 G/cm in the magnetic trap, the number of atoms in the magnetic trap was $\sim$ 1 $\times$ $10^{7}$ (initially) and temperature was $\sim$ 240 $\mu$K.

Since the temperature of atom cloud in the magnetic trap was higher than the trap depth of rf-dressed potential, the atom cloud in the magnetic trap was evaporatively cooled to reduce the temperature. For this, a single loop antenna (diameter of 25 mm) has been used for rf induced evaporation. The rf source was connected to this antenna and the rf-field frequency was scanned from 15 MHz to 2 MHz in 2 seconds, when atoms were placed in the magnetic trap. The rf evaporation was initiated after 50 ms duration of reaching the final value of gradient in magnetic trap. This time duration is good enough for untrapped atoms to escape from the magnetic trap atom cloud. During the rf evaporation process, the number of atoms and temperature, both are reduced in the atom cloud. At the end of the evaporation sequence, we observed that atom cloud in magnetic trap had $\sim$ 6$\times$ $10^{5}$ number of atoms at temperature of $\sim$ 30 $\mu$K. With these parameters in the atom cloud, we started loading atoms in rf-dressed trap. For doing this, the evaporatively cooled atom cloud in the magnetic trap was exposed to a strong rf-field. A different 20-turn coil of diameter $\sim$80 mm (shown in Fig. \ref{fig:Set_Up_Drawing}) was used to generate the necessary rf-field for rf-dressing. The rf signal from the direct digital synthesizer (DDS) (Tektronix, AFG3102C) has been fed to the rf-amplifier (Hubert, model no: A 1020-15-200) to amplify the rf-signal. The rf-amplifier output was finally connected to the 20-turn coil for generating the dressing rf-field. 

Finally, we detect the rf-dressed atom cloud in magnetic trap by a 2f-imaging setup. For this, a weak absorption probe beam is passed though the atom cloud and the transmitted probe was imaged on a CCD camera (PCO, pixelfly). All the experimental sequences, starting from switching-on the MOT beams (through AOMs) and MOT loading, molasses formation, optical pumping, magnetic trapping, rf-evaporation, rf-dressing, and detection of atoms in rf-dressed trap were controlled by a field programmable gate array (FPGA) based c-RIO (compact-Reconfigurable Input Output Module) controller of National Instruments (NI c-RIO 9035). This controller generates the analog and digital pulses to control various instrument used in the controlled operation. The controlled instruments include AOMs, power supplies of coils, shutters , DDS, CCD camera, etc.
\section{\label{sec:Results}Results and discussion}

\begin{figure}[h!]
\centering
\advance\leftskip -0.5cm
\includegraphics[width=0.85\textwidth, height = 0.5\textwidth]{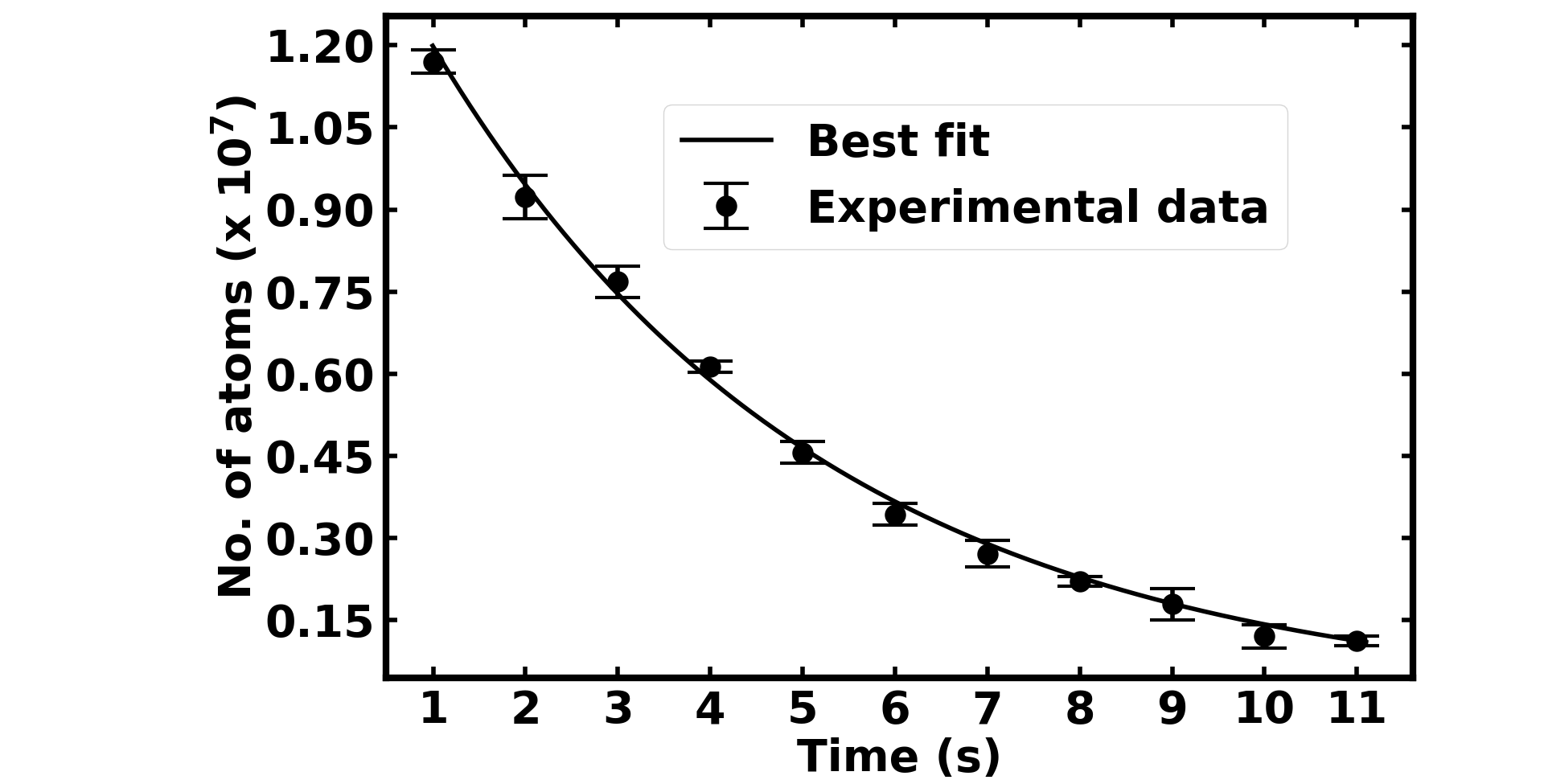}
\caption{\label{fig:Mag_Life} The measured variation (circles) in number of atoms in the quadrupole magnetic trap with  holding time in the trap. The magnetic trap was operated with final gradient of $B_{q} = 126$ G/cm. The continuous curve is an exponential fit to the measured data.}
\end{figure}

\begin{figure}[h!]
\centering
\includegraphics[width= 0.85\textwidth, height = 0.5\textheight]{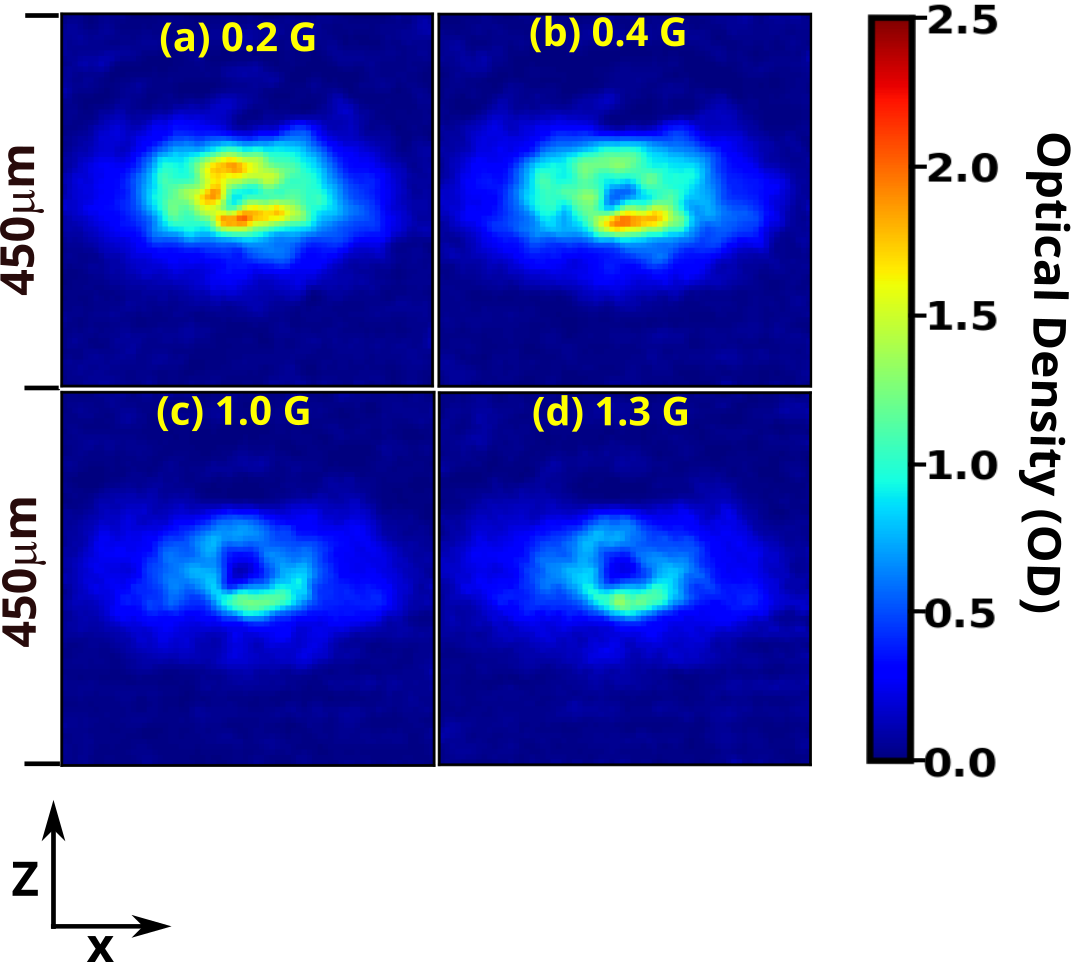}
\centering
\caption{\label{fig:Ring_Images} Experimentally observed absorption probe images (in $x-z$ plane) of $^{87}Rb$ atom cloud in rf-dressed potential for different values of dressing rf-field amplitude ($B_{rf}^{x}$). The magnetic field gradient $B_{q} = 126$ G/cm and rf-field frequency $\omega_{rf}$ $= 2\pi$ $\times$ $1.08$ MHz are common for all images from (a) to (d).}
\end{figure}

We have first characterized our magnetic trap, which is a quadrupole magnetic trap loaded in the single UHV (pressure $\sim$1.5 $\times$ $10^{-10}$ Torr) chamber setup. The number of atoms in the trap and the trap life-time were measured by analyzing the absorption probe images of the atom cloud captured at different points of time after achieving the final gradient in the magnetic trap. The variation in number atoms in the trap with time is shown in Fig. \ref{fig:Mag_Life}. For each of these measurements in Fig. \ref{fig:Mag_Life}, once the image was captured for a given data point, the cloud image for next data point was captured by running a new cycle of MOT to magnetic trap loading with changed time for camera trigger after magnetic trap loading. The error bar shows the deviation in the measured values in repeated measurements for the same point. As shown in Fig. \ref{fig:Mag_Life}, the life-time ($\frac{1}{e}$) of the magnetic trap from the best fit was $\sim 4$ s. 

Since temperature of the atom cloud in the magnetic trap ($\sim $250 $\mu$K) was higher as required for loading the rf-dressed trap, the magnetically trapped atoms were cooled further by rf-evaporation. The rf-evaporation was performed for 2 seconds with a frequency scan from 15 MHz to 2 MHz. After evaporation, the temperature of the atom cloud was estimated to be $\sim$30 $\mu$K with $\sim$6 $\times$ $10^{5}$ number of atoms in the trap. In the magnetic trap, without rf-evaporation, $\sim$ 1$\times$$10^{7}$ atoms survive after 2 seconds duration in the magnetic trap.

After evaporative cooling and reduction in temperature, atoms in magnetic trap were exposed to a strong rf-field for trapping them in rf-dressed potential. For this we have used a 20-turn circular loop antenna placed at distance of $\sim$6 cm from the atom cloud. The access of the coil was along $x$-axis (Fig. \ref{fig:Set_Up_Drawing}). This coil was driven by the rf-source (DDS) at 1.08 MHz frequency. The source signal was amplified and fed to the coil through impedance matched circuits. After exposure to the dressing rf-field for $\sim$200 ms duration, the cloud images were captured by absorption probe imaging technique. Fig. \ref{fig:Ring_Images} shows the observed absorption images of atom cloud trapped in rf-dressed potential for different values of rf-field amplitude. From these 2D images, we note that the center of atom trap has low number density of atoms. More atoms are distributed in the peripheral region of the 2D-image, which is an indicative of trapping of atoms on a shell surface in 3D. In order to understand these results, we have theoretically simulated the rf-dressed potential for a linearly polarized rf-field which are discussed as follows.

\begin{figure}[h!]
\centering
\advance\leftskip -0.45cm
\includegraphics[width=1.\textwidth, height = 0.3\textwidth]{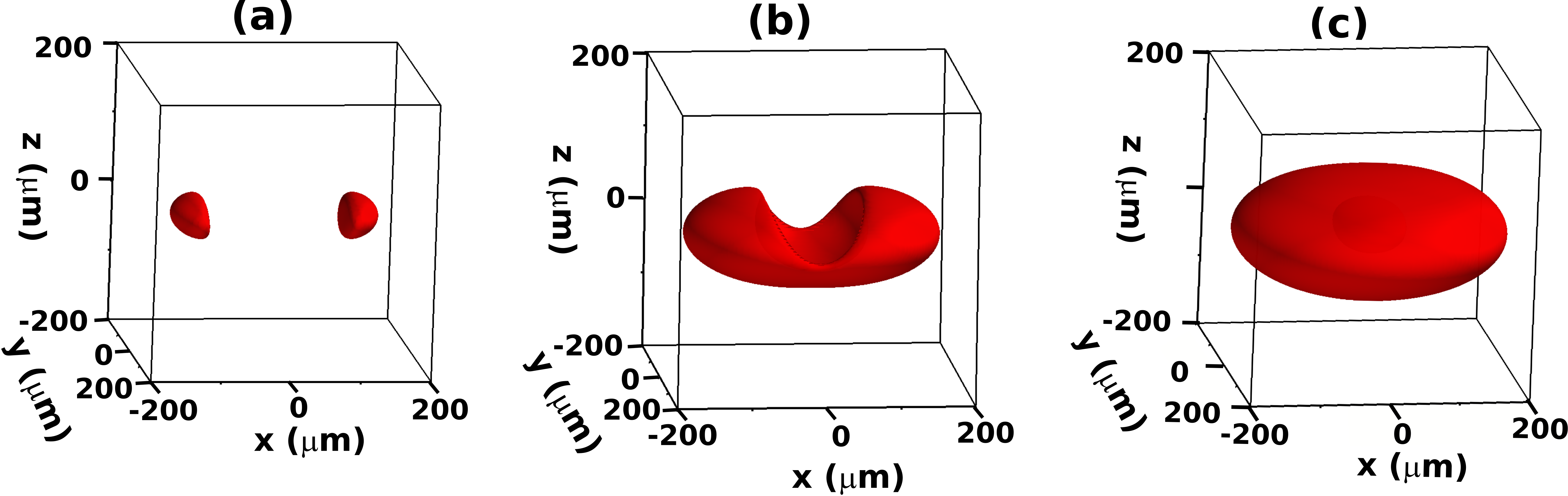}
\caption{\label{fig:3D_Potential} The 3D-plots ((a)-(c)) of rf-dressed potential $V_{D}$ for a $x$-directional linearly polarized rf-field. Each plot shows $V_{D}$ values from minimum ($\sim$1 $\mu$K) to a maximum $V_{D}$ value. The maximum value of $V_{D}$ was set as 20 $\mu$K (a), 40 $\mu$K (b), and 60 $\mu$K (c) above the minimum value. The rf-field magnitude is $B_{rf}^{x} = 1.3 G$ and frequency is $\omega_{rf} = 2\pi \times 1.08$ MHz. The value of magnetic field gradient is $B_{q}= 126$ G/cm.}
\end{figure}

Fig. \ref{fig:3D_Potential} shows a 3D-plots of contours of rf-dressed potential given by Eq. (\ref{Dressed_Potential}) for a $x$-directional linearly polarized rf-field configuration. Each plot has contours of $V_{D}$ from a minimum value ($\sim$1 $\mu$K) of $V_{D}$ to a maximum value of $V_{D}$. Here maximum $V_{D}$ value contour lies on the surface of the 3D plot. In each plot ((a), (b) and (c)), the maximum value of $V_{D}$ was respectively taken as 20 $\mu$K, 40 $\mu$K, and 60 $\mu$K above the minimum value. From these plots, one can understand the 3D configuration of trapping geometry, if kinetic energy (temperature) of atoms is known. For example, from the plot (a) in Fig. \ref{fig:3D_Potential}, one can predict that for atoms with kinetic energy smaller than 20 $\mu$K, the trapping geometry is expected to be a double-well oriented along $x$-direction. This is as predicted earlier also \cite{Lesanovsky_PRA_2006,Chakraborty2014}. 

\begin{figure}[h!]
\centering
\includegraphics[width= 0.9985\textwidth, height = 0.25\textwidth]{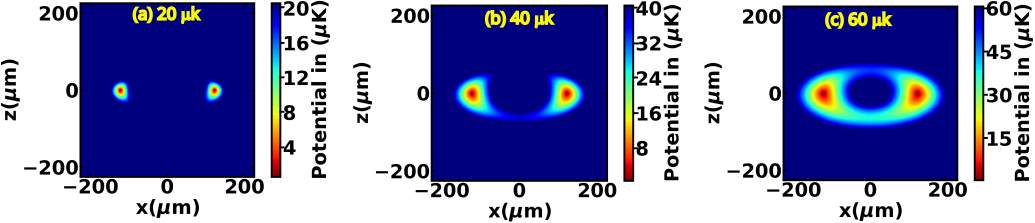}
\centering
\caption{\label{fig:2D_Potential_Contour}
Two dimensional ($x-z$) contour plots of rf-dressed potentials for different maximum values of $V_{D}$. The parameters are : $B_{q} = 126$ G/cm, rf-field frequency $\omega_{rf} = 2\pi\times$ 1.08 MHz, and amplitude of $x$-polarized rf-field as $B_{rf}^{x} = 1.3$ G. The contour (a) shows the range of the trapping potential from minimum value to 20 $\mu$K above the minimum value. Accordingly the contours in (b) and (c) show the maximum value of the trapping potential to be 40 $\mu$K and 60 $\mu$K above the minimum value.}
\end{figure}

However, for atoms with kinetic energy higher than 20 $\mu$K, the atom trapping geometry will correspond to potential contours shown in plots (b) and (c) in Fig. \ref{fig:3D_Potential}, which will be a hollow shell type atom trap configuration and not a double-well.\\

To understand it more clearly, we have also plotted the 2D contours ($x$-$z$ plane) of trapping potential $V_{D}$ with the color coding, and results are shown in Fig. \ref{fig:2D_Potential_Contour}. Here ($x-z$) contours of potential $V_{D}$ have been plotted for different maximum values of $V_{D}$. This figure shows that for the lower values of $V_{D}$ (e.g. plot (a) in Fig. \ref{fig:2D_Potential_Contour}), the trapping potential is a double-well. However, at higher values of $V_{D}$ (e.g. plot (c)), the trapping potential takes a shape of ring structure in the $x-z$ plane, which is corresponding to be a shell trap in 3D. In the experiments, we have observed (Fig. \ref{fig:Ring_Images}) hollow ellipsoidal shell trap in which an atom cloud of 30 $\mu$K temperature has been used. The atom cloud of 30 $\mu$K temperature will have significant fraction of atoms with kinetic energy more than 30 $\mu$K. Such atoms are expected to follow the trap potential shapes as shown in plots (b) and (c) in Fig. \ref{fig:2D_Potential_Contour} and plots (b) and (c) in Fig. \ref{fig:3D_Potential}, which is corresponding to a trap geometry of shell rather than a double-well. This explains our experimental observation of a shell trap for $^{87}Rb$ atoms obtained after rf-dressing with linearly polarized rf-field.

\section{\label{sec:Conclusions}Conclusion}
We have investigated the atom trapping geometry formed by superposing a linearly polarized radio-frequency (rf) field on a quadrupole magnetic trap. The hollow ellipsoidal shell type atom trap has been experimentally observed, which is explained by modelling theoretically the rf-dressed potentials. This type of trap geometry may be useful in the study of low-dimensional Bose-Einstein condensation (BEC), superfluidity, atom gyroscope, etc.

\begin{acknowledgments}
We thank S. Singh and V. Singh for useful technical discussions. We are also thankful to Shri D. Sharma and Shri Bhupinder Singh for the development of switching circuit, Shri A. Pathak and Smt S. Tiwari for the development of c-RIO based electronic controller, and P. S. Bagduwal for development of rf-amplifier and rf-impedance matching circuits. S. Sarkar is grateful for financial support from HBNI, RRCAT.
\end{acknowledgments}

\section*{Data Availability Statement}
Data available on request from the authors.




\end{document}